\documentclass[12pt,twoside]{article}
\usepackage{fleqn,espcrc1,epsfig}
\usepackage{graphicx}

\title{Electromagnetic $K^+$ production on the deuteron with
       hyperon recoil polarization}
\author{K. Miyagawa\,\address[OUS]{Department of Applied Physics,
 Okayama University of Science, 1-1 Ridai-cho, Okayama 700, Japan},
 H. Yamamura\,\addressmark  , T. Mart\,\address{Jurusan Fisika, FMIPA,
Universitas Indonesia, Depok 16424, Indonesia},
C.~Bennhold\,\address[GW]{Center for Nuclear Studies, The George
Washington University, Washington, D.C. 20052}, H.
Haberzettl\,\addressmark \, and W.~Gl\"ockle\,\address{Institut
f\"ur Theoretische Physik II, Ruhr-Universit\"at Bochum, D-44780
Bochum, Germany} }

\begin{document}

\maketitle

\begin{abstract}
Photo- and electroproduction processes of $K^+$ on the deuteron
are investigated theoretically. Modern hyperon-nucleon forces as
well as an updated kaon production operator on the nucleon are
used. Sizable
effects of the hyperon-nucleon final state interaction are seen in
various observables. Especially the photoproduction double
polarization observable $C_z$ is shown to provide a handle to
distinguish different hyperon-nucleon force models.
\end{abstract}

\section{INTRODUCTION}
Recent rigorous calculations of light hypernuclei \cite{mihi} have
contributed interesting insight into low-energy properties of the $YN$
interaction above the $\Lambda$ threshold. However, no clear
understanding of the $YN$ interaction has emerged around the
$\Sigma$ threshold. Electro- and photoproduction processes of
$K^+$  on light nuclei offer a unique possibility for studying the
$YN$ interaction in the continuum, especially near the $\Sigma$
threshold. An inclusive $d(e, e'K^+)YN$ experiment has already
been performed, and the data for $d(\gamma, K^+ Y)N$ and
$^3$He$(\gamma, K^+ Y)N$ are being analyzed at TJNAF.

We have analyzed the inclusive $d(\gamma, K^+)$  and exclusive
$d(\gamma, K^+ \Lambda (\Sigma ))$ processes \cite{gamd}, and
report here preliminary results of the electroproduction process
$d(e, e'\, K^+)$.  This study aims to investigate the coupled
$\Lambda N-\Sigma N$ interaction in the final state and
incorporates the modern $YN$ interactions of the Nijmegen group,
NSC97f \cite{nij}  and NSC89 which have been found to give a
reasonable binding energy for the hypertriton. Kaon
photoproduction on the deuteron is also important since it allows
access to the elementary cross sections on the neutron, such as
$\gamma + n \rightarrow K^+ + \Sigma^-$, in kinematic regions
where final-state interaction effects are small.

\section{PHOTOPRODUCTION}

\begin{figure}[hb!]
\begin{center}
\epsfig{file=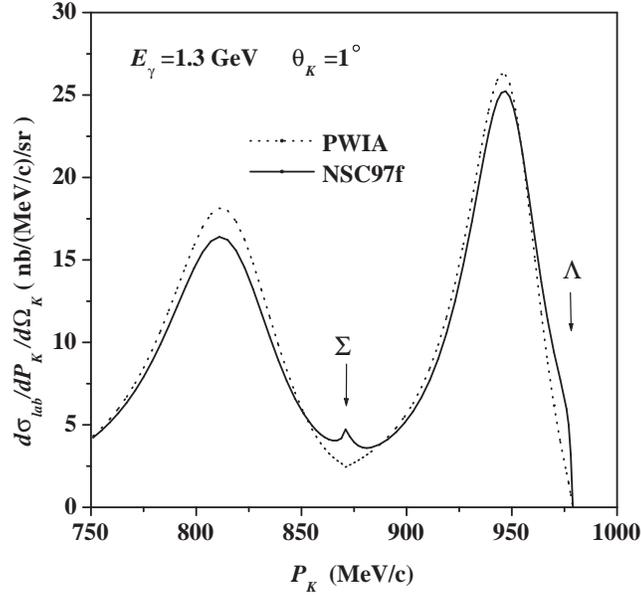,height=80mm}
\end{center}
\vspace{-10mm} \caption{Inclusive $d(\gamma ,K^+)$ cross section
as a function of kaon lab momentum $P_K$. The $K^+\Lambda N$ and
$K^+\Sigma N$ thresholds are indicated by the arrows.}
\end{figure}

%%%%%%%%%%%%%%%%%%%%%%%%%%%

\begin{figure}[hb!]
\begin{minipage}[t]{80mm}
\epsfig{file=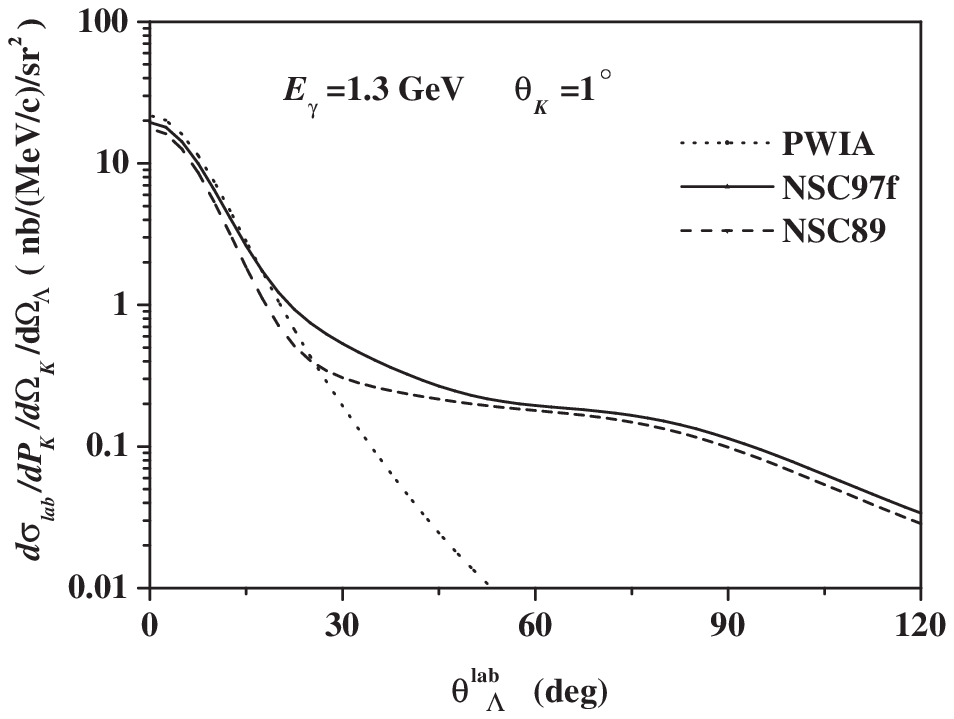, width=80mm} \caption{Exclusive
$d(\gamma,K^+\Lambda )$ cross section for lab momentum $P_K=870$
MeV/c.}
\end{minipage}
\hspace{\fill}
\begin{minipage}[t]{75mm}
\epsfig{file=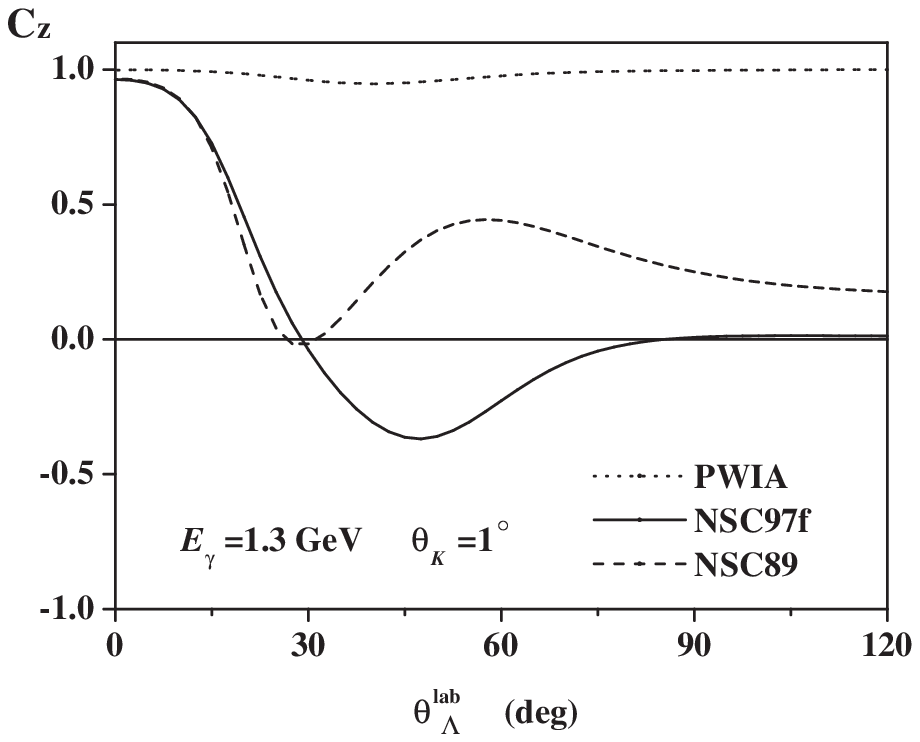,width=75mm} \caption{Double polarization
observable $C_z$  for the reaction $d(\gamma ,K^+\Lambda )$ at lab
momentum $P_K=870$ MeV/c.}
\end{minipage}
\end{figure}

%%%%%%%%%%%%%%%%%%%%%%%%%%%%%%%%%

\begin{figure}[h!]
\begin{minipage}[t]{80mm}
\epsfig{file=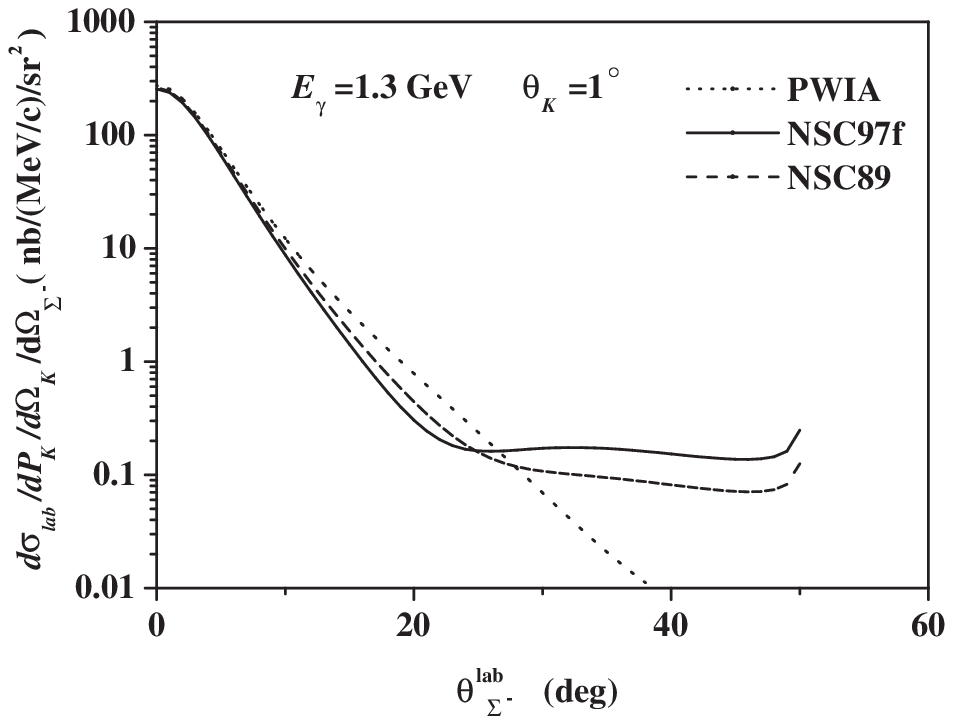, width=80mm} \caption{Exclusive $d(\gamma
,K^+\Sigma ^- )$ cross section for lab momentum $P_K=810$ MeV/c.}
\end{minipage}
\hspace{\fill}
\begin{minipage}[t]{75mm}
\epsfig{file=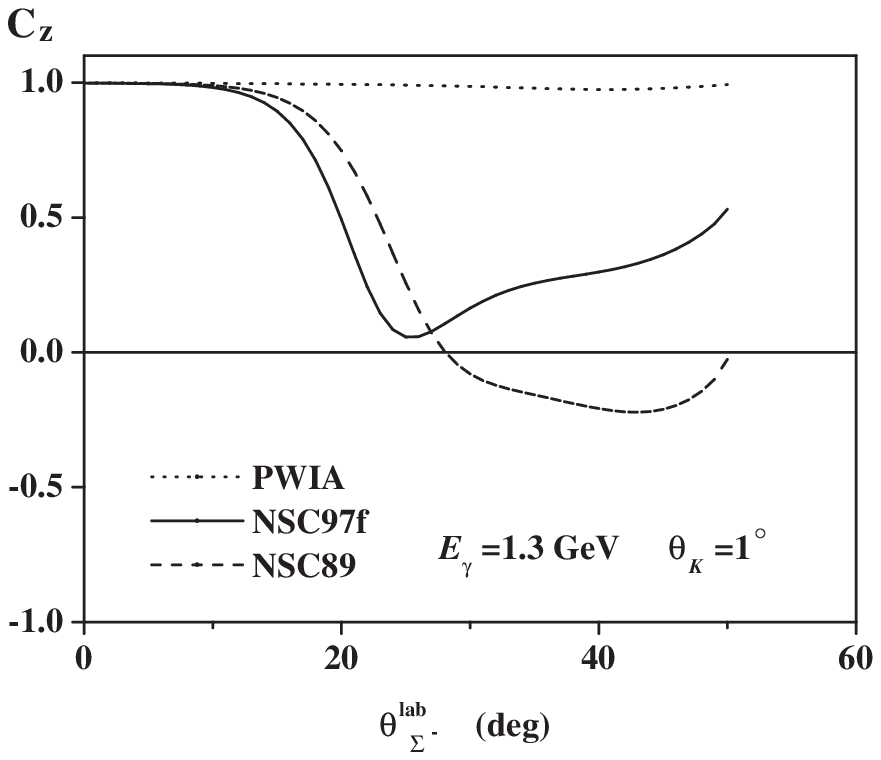,width=75mm} \caption{Double polarization
observable $C_z$ for the reaction $d(\gamma ,K^+\Sigma ^- )$ at
lab momentum $P_K=810$ MeV/c.}
\end{minipage}
\end{figure}

%%%%%%%%%%%%%%%%%%%%%%%%%%%%%%%%%

Numerical results of the inclusive $d(\gamma, K^+)$ cross
sections, using an updated production operator \cite{oper}, are
shown as a function of lab momentum $P_K$ in Fig. 1. For the
details of our theoretical formulation, we refer the reader to
Ref. \cite{gamd}. The incident photon energy is 1.3 GeV, while the
outgoing kaon angle is fixed to 1 degree. The two pronounced peaks
around $P_K=945$ and $809$ MeV/c are due to the quasifree
scattering between photon and one of the nucleons. The results
with the final state $YN$ interaction NSC97f are compared to the
PWIA results. Sizable FSI effects are seen around both $\Lambda$
and $\Sigma$ thresholds, and to a lesser degree at the two
quasifree peak positions.

For the same $E_\gamma$ and $\theta_K$, the exclusive $d(\gamma,
K^+ \Lambda)n$ cross section and double polarization observable
$C_z$ at $P_K=870$ MeV/c are shown in Figs. 2 and 3, respectively.
Figures 4 and 5 depict these observables for  $d(\gamma, K^+
\Sigma^-)p$ at $P_K=810$ MeV/c. As indicated in Fig. 1, the former
value of $P_K$ is close to the $K^+ \Sigma N$ threshold, while the
latter one corresponds to the $\Sigma$ quasifree peak position.
While the values for $C_z$ in PWIA are almost 100\%, the FSI
results show dramatic deviations.  Furthermore, the two $YN$
forces of NSC97f and NSC89 become clearly distinguishable for this
observable. Experimentally, measuring this observable involves
using circularly polarized photons along with detecting the recoil
polarization of the hyperon in the final state.

\section{ELECTROPRODUCTION}

\begin{figure}[t]
\begin{minipage}[t]{80mm}
\epsfig{file=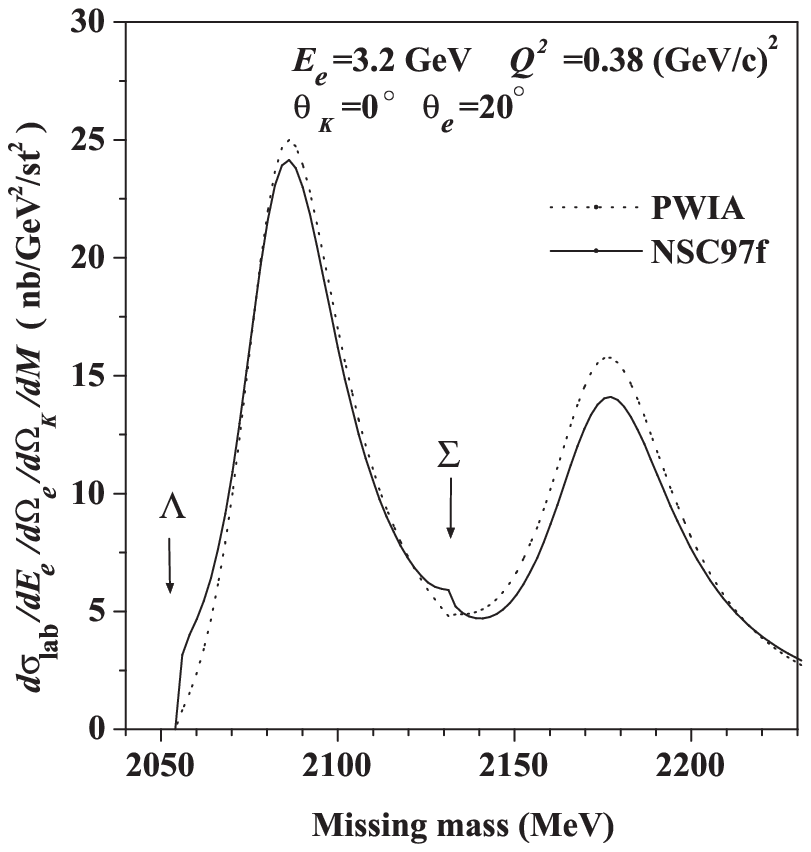,height=80mm} \caption{Missing mass spectrum
for the reaction $d(e,e'K^+)$. Results with the $YN$ final state
interaction NSC97f are compared to PWIA results.}
\end{minipage}
\hspace{\fill}
\begin{minipage}[t]{75mm}
\epsfig{file=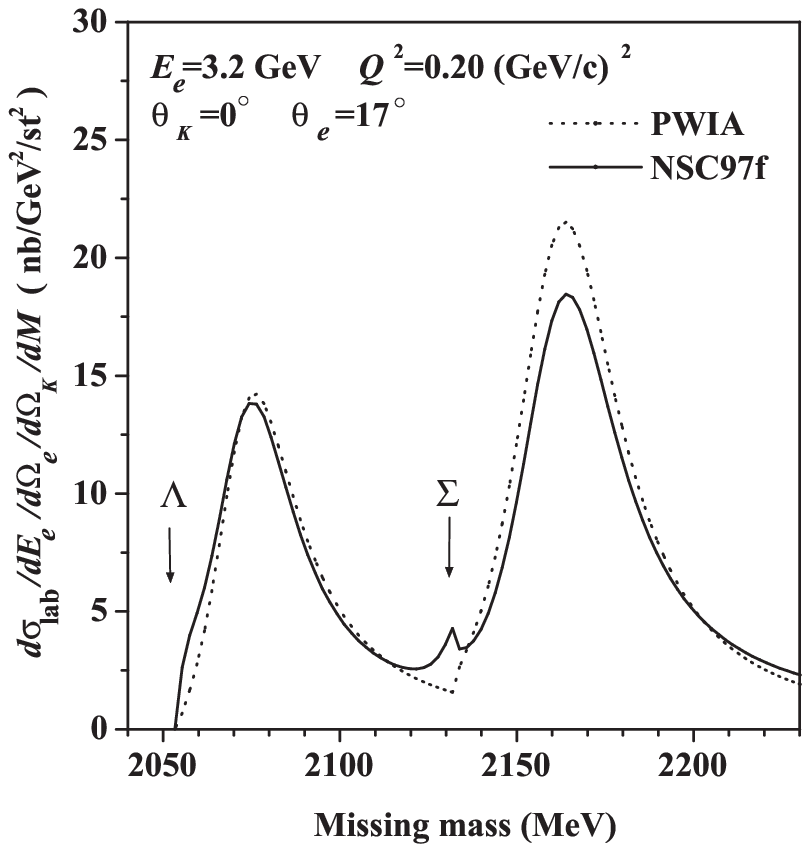,height=80mm} \caption{Same as Fig. 6, but
for $Q^2=0.20$ (GeV/c)$^2$, $\theta_e =17^{\circ}$.}
\end{minipage}
\vspace{-5mm}
\end{figure}

%%%%%%%%%%%%%%%%%%%%%%%%%%%%%%%%%%%%%%

Here we present first preliminary results for the
electroproduction process  $d(e,e'K^+)$. Two sets of results are
shown in Figs. 6 and 7. The incident electron energy $E_e$ and
momentum transfer $Q^2$ in Fig. 6 is set to reproduce the
conditions of the recent Hall C experiment at TJLAB \cite{exper}.
As in the case of the photoproduction $d(\gamma ,K^+)$, $YN$ FSI
effects are seen near both $\Lambda$ and $\Sigma$ threshold.
However, in Fig. 6, the FSI has effects in a wide range above the
$\Sigma$ threshold.  Figure 7 shows a prominent enhancement around
the $\Sigma$ threshold which is not a simple threshold effect but
is caused by a $YN$ $t$-matrix pole in the complex momentum plane
\cite{pole}.

\section{OUTLOOK}
We investigate cross sections and hyperon polarization for the
$K^+$ photoproduction on the deuteron, and find large
hyperon-nucleon FSI effects in the double polarization observable
$C_z$. Also, in the electroproduction, for suitable $Q^2$ values,
cross sections show a prominent enhancement around the $\Sigma$
threshold. A systematic analysis for a wide range of kinematics
for both photo- and electroproduction processes is in progress.
Future studies will investigate final-state interaction effects in
kaon photo- and electroproduction on the $A=3$ system.

\end{document}